\documentclass[aps,tightenlines,nofootinbib,11pt]{revtex4}

\usepackage{graphicx}
\usepackage{amsfonts}
\usepackage{amsmath}
\usepackage{amsthm}
\usepackage{amssymb}
\usepackage{setspace}
\usepackage{epsfig}

\usepackage{hyperref}      
\newcommand{\beq}{\begin{equation}} 
\newcommand{\eeq}{\end{equation}}
\newcommand{\met}{\rlap{\,\,/}E_T}

\begin{document}

\title{Kaluza-Klein Dark Matter: Direct Detection vis-a-vis LHC}

\author{Sebastian Arrenberg} \affiliation{Department of Physics, University of Z\"urich, Z\"urich, 8057, Switzerland}
\author{Laura Baudis} \affiliation{Department of Physics, University of Z\"urich, Z\"urich, 8057, Switzerland} 
\author{Kyoungchul Kong} \affiliation{Department of Physics and Astronomy, University of Kansas, Lawrence, KS 66045 USA} 
\author{Konstantin T. Matchev} \affiliation{Institute for Fundamental Theory, Physics Department, University of Florida, Gainesville, FL 32611, USA}
\author{Jonghee Yoo} \affiliation{Fermi National Accelerator Laboratory, Batavia, IL 60510, USA} 

\noaffiliation

\begin{abstract}
We present updated results on the complementarity between high-energy colliders and dark matter direct detection 
experiments in the context of Universal Extra Dimensions (UED). 
In models with relatively small mass splittings between the dark matter candidate and the rest of the (colored) spectrum, 
the collider sensitivity is diminished, but direct detection rates are enhanced.
UED provide a natural framework to study such mass degeneracies. 
We discuss the detection prospects for the KK photon $\gamma_1$ and the KK $Z$-boson $Z_1$, 
combining the expected LHC reach with cosmological constraints from WMAP/Planck, 
and the sensitivity of current or planned direct detection experiments.
Allowing for general mass splittings, neither colliders, nor direct detection experiments by themselves 
can explore all of the relevant KK dark matter parameter space. 
Nevertheless, they probe different parameter space regions, and the combination of the two types of 
constraints can be quite powerful. 

\end{abstract}

\maketitle

We present updated results on the complementarity between high-energy colliders and 
dark matter direct detection experiments \cite{Arrenberg:2008wy}
in the context of Universal Extra Dimensions \cite{Appelquist:2000nn}.
As our reference, we take the mass spectrum in Minimal Universal Extra Dimensions (MUED), 
which is fixed by the radius ($R$) of the extra dimension and the cut-off scale 
($\Lambda$) \cite{Cheng:2002ab,Cheng:2002iz}.
To illustrate the complementary between dark matter detection and searches at the LHC, 
we introduce a slope in the MUED mass spectrum, in terms of the mass splitting ($\Delta_{q_1}$) between 
the mass of the lightest Kaluza-Klein (KK) partner (LKP) $m_{LKP}$ and the KK quark mass $m_{q_1}$:
$$\Delta_{q_1} = \frac{m_{q_1} - m_{LKP}}{m_{LKP}}.$$ 
We take $\Delta_{q_1}$ as a free parameter, which is possible in a more general framework 
with boundary terms and bulk masses (see, e.g., \cite{Flacke:2013pla}).
The LKP is taken to be either the KK mode $\gamma_1$ of the photon (as in MUED),
or the KK mode $Z_1$ of the $Z$-boson. In the latter case, we assume that the gluon and the 
remaining particles to be respectively 20\% and 10\% heavier than the $Z_1$. 
This choice is only made for definiteness, and does not impact our results, 
as long as the remaining particles are sufficiently heavy and do not participate in co-annihilation processes.

\begin{figure}[t]
\includegraphics[width=0.475\textwidth]{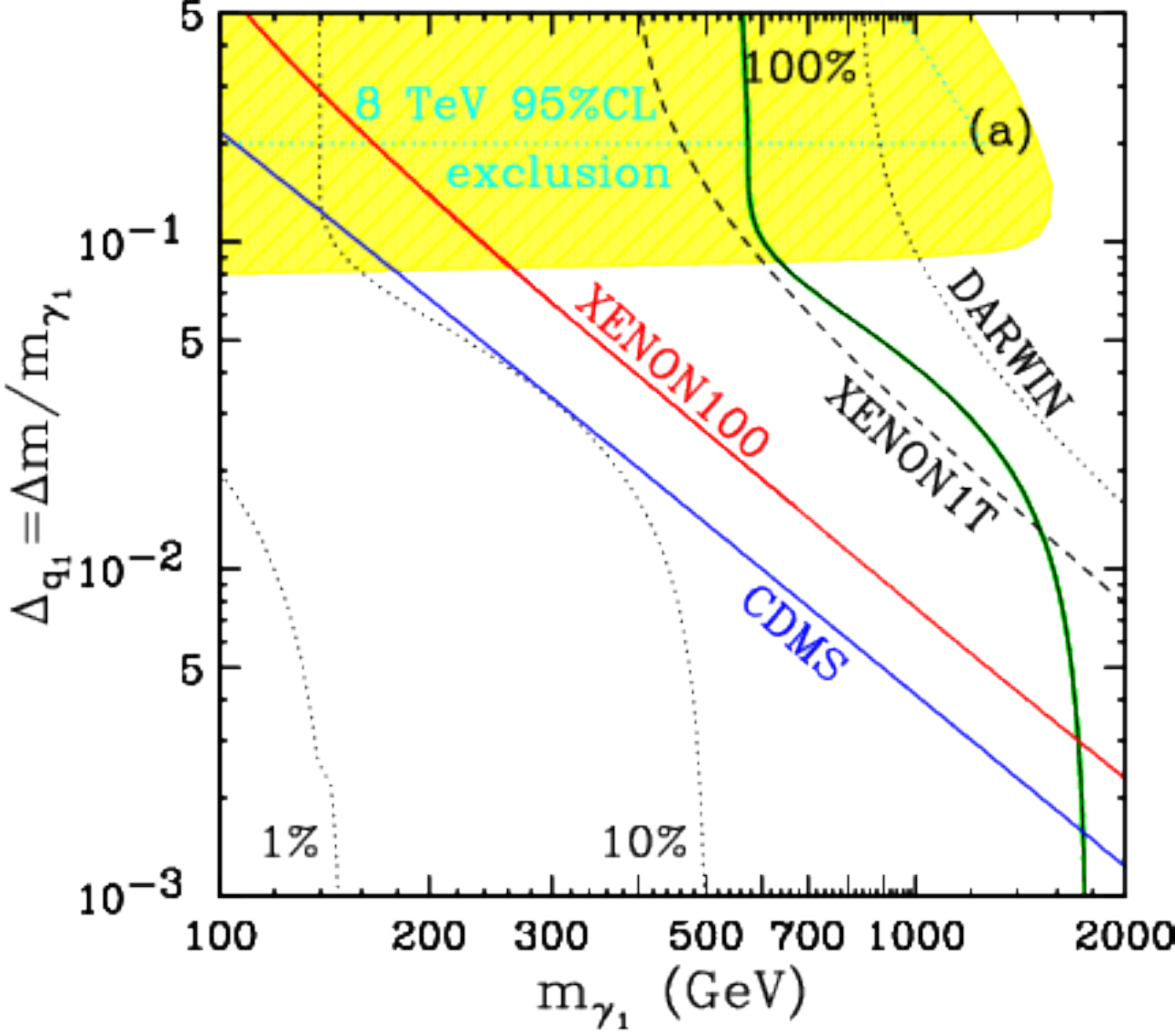}
\hspace{0.1cm}
\includegraphics[width=0.475\textwidth]{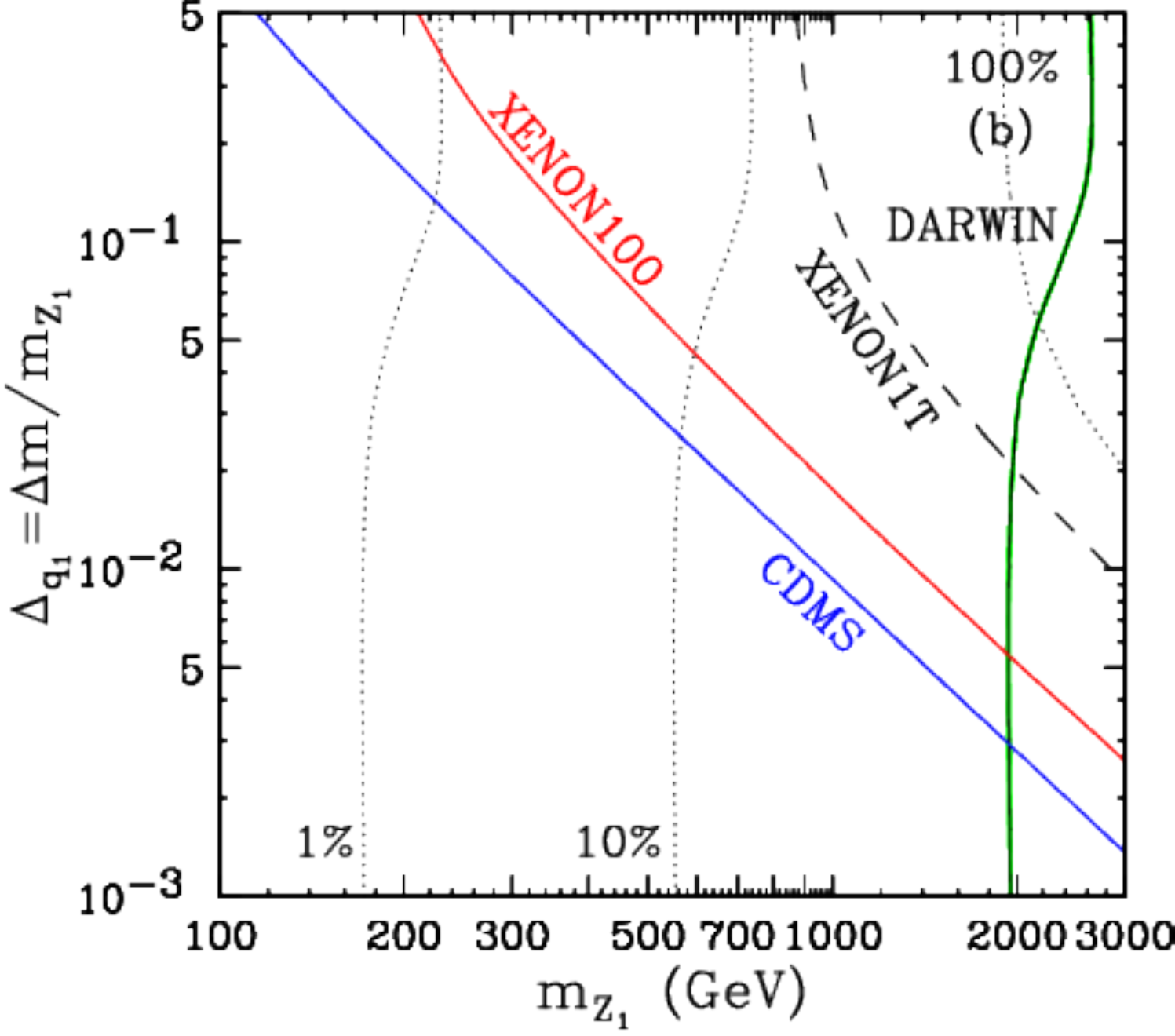}
\vspace{-0.2cm}
\caption{\sl
Combined plot of the direct detection limit on the spin-independent cross section, 
the limit from the relic abundance and the LHC reach for (a) $\gamma_1$ and (b) $Z_1$, 
in the parameter plane of the LKP mass and the mass splitting $\Delta_{q_1}$. 
The remaining KK masses have been fixed as in Ref. \cite{Cheng:2002iz} 
and the SM Higgs mass is $m_h=125$\,GeV. $\Lambda R=20$ is assumed. 
The black solid line accounts for all of the dark matter (100\%) 
and the two black dotted lines show 10\% and 1\%, respectively. 
The green band shows the WMAP/Planck range, $0.117 < \Omega_{CDM}h^2 < 0.1204$.
The blue (red) solid line labelled by CDMS (XENON100) 
shows the current limit of the experiment whereas the dashed and dotted lines 
represent projected limits of future experiments. 
In the case of $\gamma_1$ LKP, a ton-scale experiment will rule out 
most of the parameter space while there is little parameter space left in the case of $Z_1$ LKP. 
The yellow region in the case of $\gamma_1$ LKP shows parameter space 
that could be covered by the collider search in the $4\ell+\met$ channel 
at the LHC with a luminosity of 100 fb$^{-1}$ \cite{Cheng:2002ab}. 
}
\label{fig:SI_Delta_Neutron_B_Z}
\end{figure}

In the so defined $(m_{LKP},\Delta_{q_1})$ parameter plane, in Fig.~\ref{fig:SI_Delta_Neutron_B_Z} we
superimpose the limit on the spin-independent elastic scattering cross section, 
the limit on the relic abundance and the LHC reach in the four 
leptons plus missing energy ($4\ell + \met$)
channel which has been studied in~\cite{Cheng:2002ab} at the 14 TeV 
(see Ref. \cite{Belyaev:2012ai} for 7+8 TeV). This signature
results from the pair production (direct or indirect) of $SU(2)_W$-doublet 
KK quarks, which subsequently decay to $Z_1$'s and jets. The leptons (electrons or muons)
arise from the $Z_1\to \ell^+\ell^-\gamma_1$ decay, whose branching fraction 
is approximately $1/3$~\cite{Cheng:2002ab}.
Requiring a 5$\sigma$ excess at a luminosity of 100 fb$^{-1}$, 
the LHC reach extends up to $R^{-1} \approx m_{\gamma_1} \sim 1.5$ TeV, 
which is shown as the right-most boundary of the (yellow) shaded region
in Fig.~\ref{fig:SI_Delta_Neutron_B_Z}a. The slope of that boundary is due to
the fact that as $\Delta_{q_1}$ increases, so do the KK quark masses, and their 
production cross sections are correspondingly getting suppressed, diminishing
the reach. We account for the loss in cross section according to the
results from Ref.~\cite{Datta:2005zs}, assuming also that, as expected, the 
level-2 KK particles are about two times heavier than those at level 1.
Points which are well inside the (yellow) shaded region, of course, 
would be discovered much earlier at the LHC. Notice, however, that the LHC reach 
in this channel completely disappears for $\Delta_{q_1}$ less than about 8\%.
This is where the KK quarks become lighter than the $Z_1$ (recall that 
in Fig.~\ref{fig:SI_Delta_Neutron_B_Z}a $m_{Z_1}$ is fixed according to
the MUED spectrum) and the $q_1\to Z_1$ decays are turned off. 
Instead, the KK quarks all decay directly to the $\gamma_1$ LKP 
and (relatively soft) jets, presenting a monumental challenge for an LHC discovery.
So far there have been no studies of the collider phenomenology of a
$Z_1$ LKP scenario, but it appears to be extremely challenging, especially if the 
KK quarks are light and decay directly to the LKP. This is why
there is no LHC reach shown in Fig.~\ref{fig:SI_Delta_Neutron_B_Z}b.
We draw attention once again to the
lack of sensitivity at small $\Delta_{q_1}$: such small mass splittings are 
quite problematic for collider searches. 
The current LHC exclusion limit (95\% C.L. at 8 TeV) on $R^{-1}$ is about 1250 GeV for $\Lambda R=20$ \cite{Belyaev:2012ai}. 
and this is shown as the dotted (cyan) line. The horizontal line at $\Delta_{q_1} \sim 0.2$ is the average mass splitting in MUED. 
To indicate roughly the approximate boundary of the excluded region, the slanted line around 1 TeV is added, 
assuming the shape of the boundary is similar to that for the LHC14 reach.

In Fig.~\ref{fig:SI_Delta_Neutron_B_Z} we contrast the LHC reach 
with the relic density constraints \cite{Servant:2002aq,Kong:2005hn} and with the sensitivity of 
direct detection experiments \cite{Cheng:2002ej,Servant:2002hb}. 
The green shaded region labelled by 100\% represents 2$\sigma$ band, 
$0.117 < \Omega_{CDM}h^2 < 0.1204$ \cite{Ade:2013zuv} and the black solid line inside this 
band is the central value $\Omega_{CDM}h^2 = 0.1187$. 
The region above and to the right of this band is disfavored
since UED would then predict too much dark matter. 
The green-shaded region is where KK dark matter 
is sufficient to explain all of the dark matter in the universe, while
in the remaining region to the left of the green band 
the LKP can make up only a fraction of the dark matter in the universe.
We have indicated with the black dotted contours the 
parameter region where the LKP would contribute only 10\% and 1\%
to the total dark matter budget. Finally, the
solid (CDMS \cite{Ahmed:2009zw} in blue and XENON100 \cite{Aprile:2012nq} in red) lines show the 
current direct detection limits, while the dotted and dashed lines 
show projected sensitivities for future experiments \cite{Aprile:2012zx,Baudis:2012bc,Baudis:2010ch}
\footnote{Here and in the rest of the paper, 
when presenting experimental limits in an under-dense or an over-dense 
parameter space region, we do not rescale the expected 
direct detection rates with the calculated relic density.
The latter is much more model-dependent, e.g. the mismatch with the relic abundance may be fixed by non-standard cosmological evolution, 
having no effect on the rest of our analysis.}.

Fig.~\ref{fig:SI_Delta_Neutron_B_Z} demonstrates the complementarity between the 
three different types of probes which we are considering. 
First, the parameter space region at very large $m_{LKP}$ is inconsistent
with cosmology -- if the dark matter WIMP is too heavy, 
its relic density is too large. The exact numerical bound on the LKP mass
may vary, depending on the particle nature of the WIMP (compare 
Fig.~\ref{fig:SI_Delta_Neutron_B_Z}a to Fig.~\ref{fig:SI_Delta_Neutron_B_Z}b)  
and the presence or absence of coannihilations (compare the
$m_{LKP}$ bound at small $\Delta_{q_1}$ to the bound at large $\Delta_{q_1}$).
Nevertheless, we can see that, in general, cosmology does provide 
an upper limit on the WIMP mass. On the other hand, colliders are 
sensitive to the region of relatively large mass splittings $\Delta_{q_1}$,
while direct detection experiments are at their best at small
$\Delta_{q_1}$ {\em and} small $m_{LKP}$. 
The relevant parameter space is therefore getting squeezed from
opposite directions and is bound to be covered eventually.
This is already seen in the case of $\gamma_1$ LKP from 
Fig.~\ref{fig:SI_Delta_Neutron_B_Z}a: the future experiments 
push up the current limit almost to the WMAP/Planck band. 
In the case of $Z_1$ LKP the available parameter space 
is larger and will not be closed with the currently envisioned experiments 
alone. However, one should keep in mind that detailed LHC studies
for that scenario are still lacking.

\begin{figure}[t]
\includegraphics[width=0.475\textwidth]{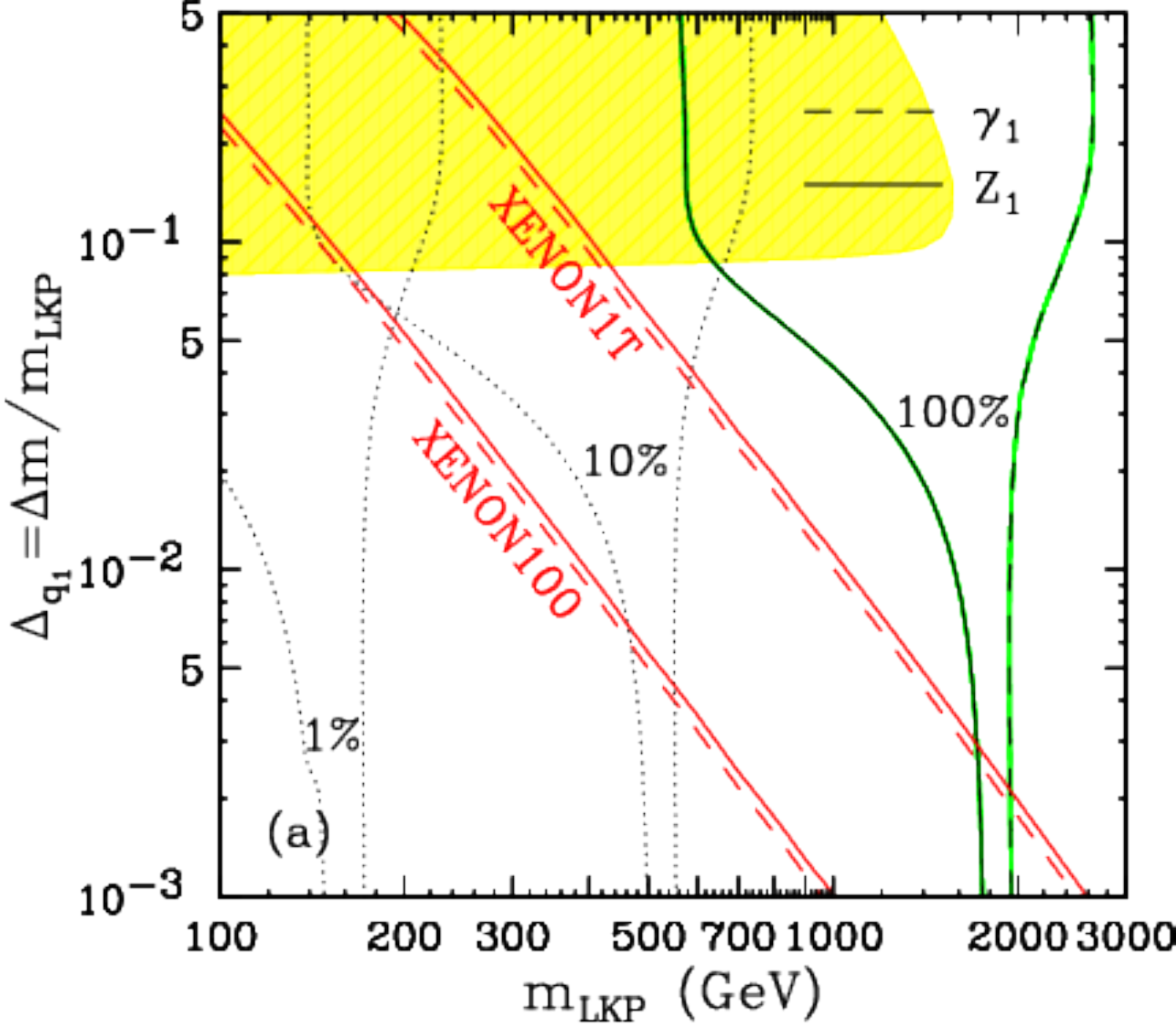}
\hspace{0.1cm}
\includegraphics[width=0.475\textwidth]{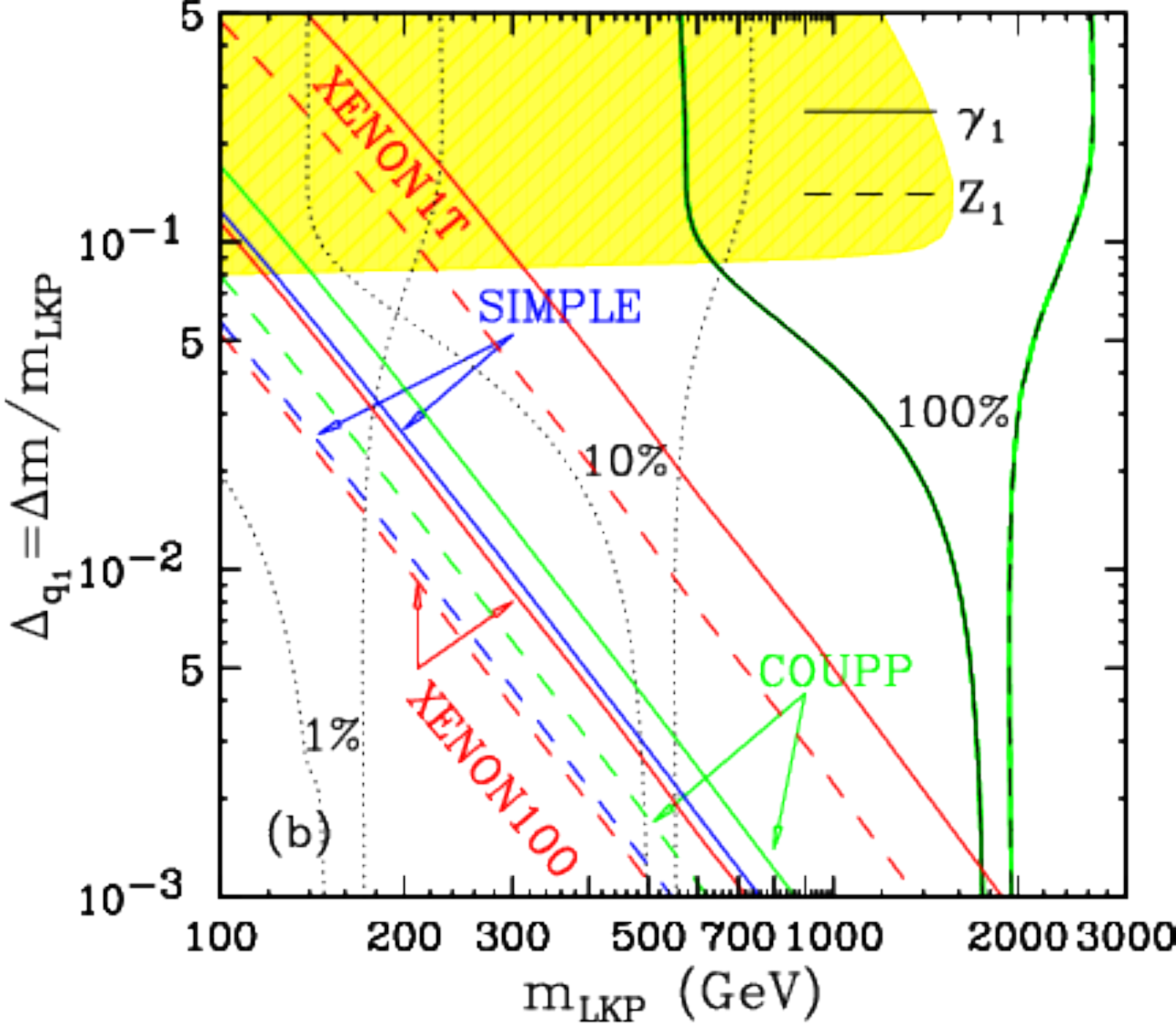}
\vspace{-0.2cm}
\caption{\sl Experimental upper bounds (90\% C.L.) on the spin-dependent 
elastic scattering cross sections on (a) neutrons and (b) protons
in the $m_{LKP}$-$\Delta_{q_1}$ plane. 
The solid (dashed) curves are limits on $\gamma_1$ ($Z_1$) from each experiment. Shaded regions and dotted lines are defined in the same way as in Fig.~\ref{fig:SI_Delta_Neutron_B_Z}. The depicted LHC reach (yellow shaded region) applies only to the case of $\gamma_1$ LKP. }
\label{fig:SD_Delta_B}
\end{figure}

Similarly the spin-dependent elastic scattering cross sections also exhibit an enhancement at small $\Delta_{q_1}$.
In Fig.~\ref{fig:SD_Delta_B}
we combine existing limits from three different experiments 
(XENON100 \cite{Aprile:2013doa}, SIMPLE \cite{Felizardo:2011uw} and COUPP \cite{Behnke:2012ys}) in the $(m_{LKP},\Delta_{q_1})$ plane.
Panel (a) (panel (b)) shows the constraints from the 
WIMP-neutron (WIMP-proton) SD cross sections. The rest 
of the KK spectrum has been fixed as in Fig. \ref{fig:SI_Delta_Neutron_B_Z}. 
The solid (dashed) curves are limits 
on $\gamma_1$ ($Z_1$) from each experiment. The constraints from 
LHC and WMAP on the $(m_{LKP},\Delta_{q_1})$ parameter space are the same as in 
Fig.~\ref{fig:SI_Delta_Neutron_B_Z}.

By comparing Figs.~\ref{fig:SI_Delta_Neutron_B_Z} and \ref{fig:SD_Delta_B}
we see that, as expected, the parameter space constraints for SI interactions 
are stronger than those for SD interactions. For example, in perhaps 
the most interesting range of LKP masses from 300\,GeV to 1 TeV, the SI limits 
on $\Delta_{q_1}$ in Fig.~\ref{fig:SI_Delta_Neutron_B_Z} range from $\sim 10^{-1}$ 
down to $\sim 10^{-2}$. On the other hand, the
SD bounds on $\Delta_{q_1}$ for the same range of $m_{LKP}$ are about an 
order of magnitude smaller (i.e. weaker). We also notice that the constraints 
for $\gamma_1$ LKP are stronger than for $Z_1$ LKP. This can be easily understood since 
for the same LKP mass and KK mass splitting, the $\gamma_1$ SD cross sections 
are typically larger.

Fig.~\ref{fig:SD_Delta_B} also reveals that the experiments rank 
differently with respect to their SD limits on protons and neutrons. For example,
SIMPLE and COUPP are more sensitive to the proton cross section, while
XENON100 is more sensitive to the neutron cross section.
As a result, the current best SD limit on protons comes from COUPP,
but the current best SD limit on neutrons comes from XENON100.

\vspace{-1.3cm}
\begin{acknowledgments}
\vspace{-0.3cm}
This work is in part supported by the Swiss National Foundation SNF and the U.S. Department of Energy.
\end{acknowledgments}
\vspace{-0.5cm}



\end{document}